\documentclass{aa}
\usepackage{natbib}
\usepackage{epsfig}
\def\ut#1{\mathop{\vtop{\ialign{##\crcr
     $\hfil\displaystyle{#1}\hfil$\crcr\noalign
     {\kern1pt\nointerlineskip}\hbox{$\hfil\sim\hfil$}\crcr
     \noalign{\kern1pt}}}}}

\def\undersymbol#1#2{\mathop{\vtop{\ialign{##\crcr
     $\hfil\displaystyle{#2}\hfil$\crcr\noalign
     {\kern1pt\nointerlineskip}\hbox{$\hfil#1\hfil$}\crcr
     \noalign{\kern1pt}}}}}
\def\arcsec{^{\prime\prime}}
\def\arcmin{^{\prime}}
\def\degr{^0}

\usepackage{graphicx}

\begin{document}

\title{Rotating baryonic dark halos}

\author{F. De Paolis\inst{1,2}, A.V. Gurzadyan\inst{3}, A.A. Nucita\inst{1,2},
        V.G. Gurzadyan\inst{4,5}, A. Qadir\inst{6}, A. Kashin\inst{4}, A. Amekhyan\inst{4}, S. Sargsyan\inst{4}, Ph. Jetzer\inst{7}, G. Ingrosso\inst{1,2} \and N. Tahir\inst{8}
}

\institute{
        Department of Mathematics and Physics ``E. De Giorgi'', University of Salento, Via per Arnesano, I-73100  Lecce, Italy     \and
        INFN, Sezione di Lecce, Via per Arnesano, I-73100 Lecce, Italy
                                 \and
                         Homerton College, University of Cambridge, Cambridge, UK
                           \and
                     Center for Cosmology and Astrophysics, Alikhanian National Laboratory and Yerevan State University, Yerevan, Armenia
                     \and
               SIA, Sapienza University of Rome, Rome, Italy
               \and
        Fellow of the Pakistan Academy of Sciences, Constitution Avenue, G-5, Islamabad, Pakistan
\and
Physik-Institut, Universit\"at
Z\"urich, Winterthurerstrasse 190, 8057 Z\"urich, Switzerland
\and
        Department of Physics, School of Natural Sciences,
National University of Sciences and Technology, Islamabad, Pakistan
}

   \offprints{F. De Paolis, \email{depaolis@le.infn.it}}
   \date{Submitted: XXX; Accepted: XXX}

\abstract{
Galactic halos are of great importance for our understanding of both the dark
matter nature and primordial non-Gaussianity in the perturbation spectrum, a
powerful discriminant of the physical mechanisms that generated the
cosmological fluctuations observed today. In this paper we analyze {\it
Planck} data towards the  galaxy M104 (Sombrero)  and find an asymmetry in
the microwave temperature which extends up to about $1 \degr$ from the
galactic center. This frequency-independent asymmetry is consistent with that  induced by
the Doppler effect due to the galactic rotation and we find a probability of less
than about $0.2\%$ that it is due to a random fluctuation of the microwave
background. In addition, {\it Planck} data indicate the relatively complex
dynamics of the  M104 galactic halo, and this appears to be in agreement with
previous studies.   In view of our previous analysis of the dark halos of
nearby galaxies, this finding  confirms the efficiency of the method used in
revealing and mapping the dark halos around relatively nearby edge-on
galaxies.}

   \authorrunning{De Paolis et al.}
   \titlerunning{Rotating baryonic dark halos}
   \maketitle
%

\section{Introduction}

As is well known,  baryons contribute about $5\%$ of our universe, but
observations show that at least $40-50\%$ of the baryons in the local
universe are undetected. The question therefore arises as to what form they
take. Indeed, most of these baryons must be in a form that is difficult to detect.
Cosmological simulations suggest that these baryons have been ejected from
galaxies into the intergalactic medium and are present in the form of a
warm-hot  medium around galaxies  (\citealt{ro1,ro2,gupta,fraser}) at
temperatures of about $10^5-10^7$ K.  A non-negligible fraction of these
baryons might also lie in a very cold form in clouds in the galactic halos.
This possibility, previously suggested in \cite{dijr,dijrAA,gs1996}, was more
recently directly evidenced by the detection of the so-called Herschel
cold clouds (see \cite{nieuwenhuizen2012} and references therein). Indeed,
observations by Herschel-SPIRE towards the Large and Small Magellanic Clouds reveal the presence of thousands of gas clouds with temperatures of about 15 K,
and it has been calculated that the full population of these clouds might
constitute a non-negligible fraction of the Galactic halo dark matter.

We also remark here that galactic halos are on the one hand the least
studied substructures of galaxies, and on the other are of particular
importance for understanding large-scale structure formation.
First, the parameters of galactic halos are sensitive to the non-Guassianity
in the primordial cosmological perturbation spectrum (\citealt{sk}), second,
halos are probes for the mysterious nature of the  dark matter
(\citealt{bertone2013,som}), and third, halos determine the properties of the
disks and the spheroidal structures of galaxies (\citealt{kr,vg}).

The primordial non-Gaussianity can be imprinted even in the profiles of the
galactic halos (\citealt{diz}), thus linking the latter to the inflationary
phase of the Universe (\citealt{an}). Halo properties can thus be probes for
the nature of perturbations; for example, even though we presently possess no
observable evidence of non-adiabatic (iso-curvature) perturbations, we have
strong constraints on their possible resistance and it was shown that during
re-ionization the iso-curvature modes are transferred to adiabatic ones
(\citealt{w}).

In a series of papers we have used the cosmic microwave background (CMB)
data, first that of WMAP and then that of {\it Planck,} to trace the halos of certain
nearby galaxies
(\citealt{depaolis2011,depaolis2014,depaolis2015,gurzadyan2015,depaolis2016,gurzadyan2018}).
The main aim was to test if the microwave data show a substantial temperature
asymmetry of one side with respect to the other about the rotation axis of
the galactic disks, as first suggested in \cite{dijqr1995} as due to the possible presence of a substantial amount of cold baryons in the galactic halos. 
Another motivation is that at the scales of galaxy clusters, 
the kinetic Sunyaev-Zeldovich (SZ) effect due to the cluster rotation is expected to induce  
temperature asymmetry in microwave data (see, e.g., \citealt{chluba2002,baxter2019}). 
The same could in principle also happen for galaxies, provided they have a very hot gas halo component.

Among the
considered objects were the M31 galaxy, the active radio galaxy Centaurus A,
M82, the largest galaxy in the M81 Group, the M33 galaxy, where we found a
substantial temperature asymmetry with respect to its minor axis up to about
$3 \degr$ from the galactic center and which correlates well with the HI
velocity field at 21 cm, at least within about $0.5 \degr$
\citep{depaolis2016}, and the M81 galaxy. In those galaxies we detected a
clear temperature asymmetry of one side with respect to the other (as predicted for a Doppler shift) with $\Delta T/T$ values of typically about $2-3\times 10^{-5}$  
and extending much further than the visible part of the galaxies (typically corresponding to peak values of  $\Delta T\simeq 60-80$ $\mu$K). 
Moreover,  this temperature asymmetry is always almost frequency independent, which 
 is an indication of an effect due to the galaxy rotation.  It was thereby shown
that our method can be applied to nearby edge-on spirals to trace the halo
bulk dynamics on rather large scales in a model-independent way. In other
words, the method is revealing the dark halos through  the regular motion of
either a cold gas component  (\citealt{dijqr1995}; for the possible modelling of
the gas clouds and their distribution in galaxies we refer to, e.g.,
\cite{pfenniger1994,fabian1994,depaolis1998,draine1998,walker1998}),
or a hot ionized gas component (for a discussion of hot ionized gas through the thermal and kinetic
Sunyaev-Zeldovich effect we refer the reader to,  e.g., \cite{lim}).

Along with  the studies of the mentioned galaxies, we note that  other
galaxies, namely M63, M64, M65,  and M66, analyzed in a similar manner,
either do not show any significant microwave temperature asymmetry (for the
first three galaxies) or the apparent asymmetry is frequency dependent (the
M66 galaxy), and therefore has to be attributed, in this specific case at least, to
a nonDoppler type signature \citep{ag}.

Here we continue those studies on the microwave mapping of dark halos,
analyzing  the {\it Planck} data on a nearly edge-on Sa type spiral galaxy
M104 known by its remarkable internal dynamics (see, e.g.,
\citealt{jardel2011}), and draw some conclusions on the baryonic dark matter
content of its halo.

The outline of the paper is the following: after a short presentation of the
M104 galaxy,  the analysis of {\it Planck} data is discussed in Section 2,
while in Section 3 we present our main conclusions.

\begin{figure}[ht]
\centering
  \includegraphics[width=0.48\textwidth]{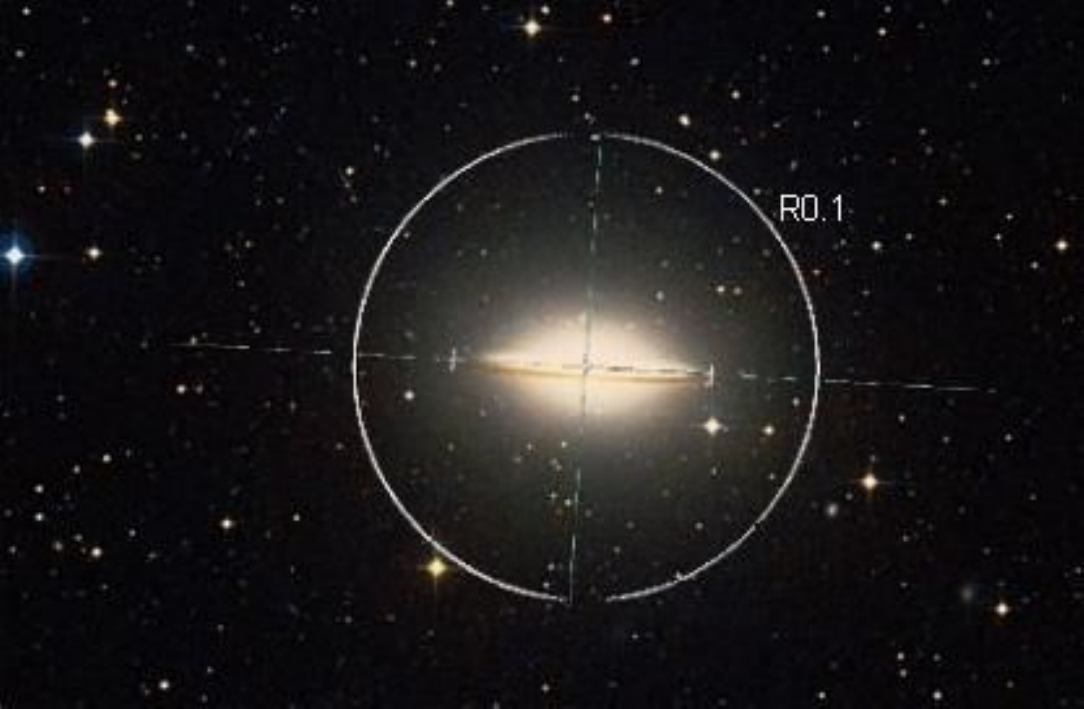}
 \caption{Sombrero galaxy in the visible band. The white circle traces the distance of $0.1\degr$ around the galaxy center at coordinates RA$: 12^{\rm h}39^{\rm m}59.4^{\rm s}$, Dec$:-11\degr 39\arcmin 23\arcsec$.} \label{FigSombrero}
 \end{figure}

\begin{figure}[ht]
\centering
\includegraphics[width=0.48\textwidth]{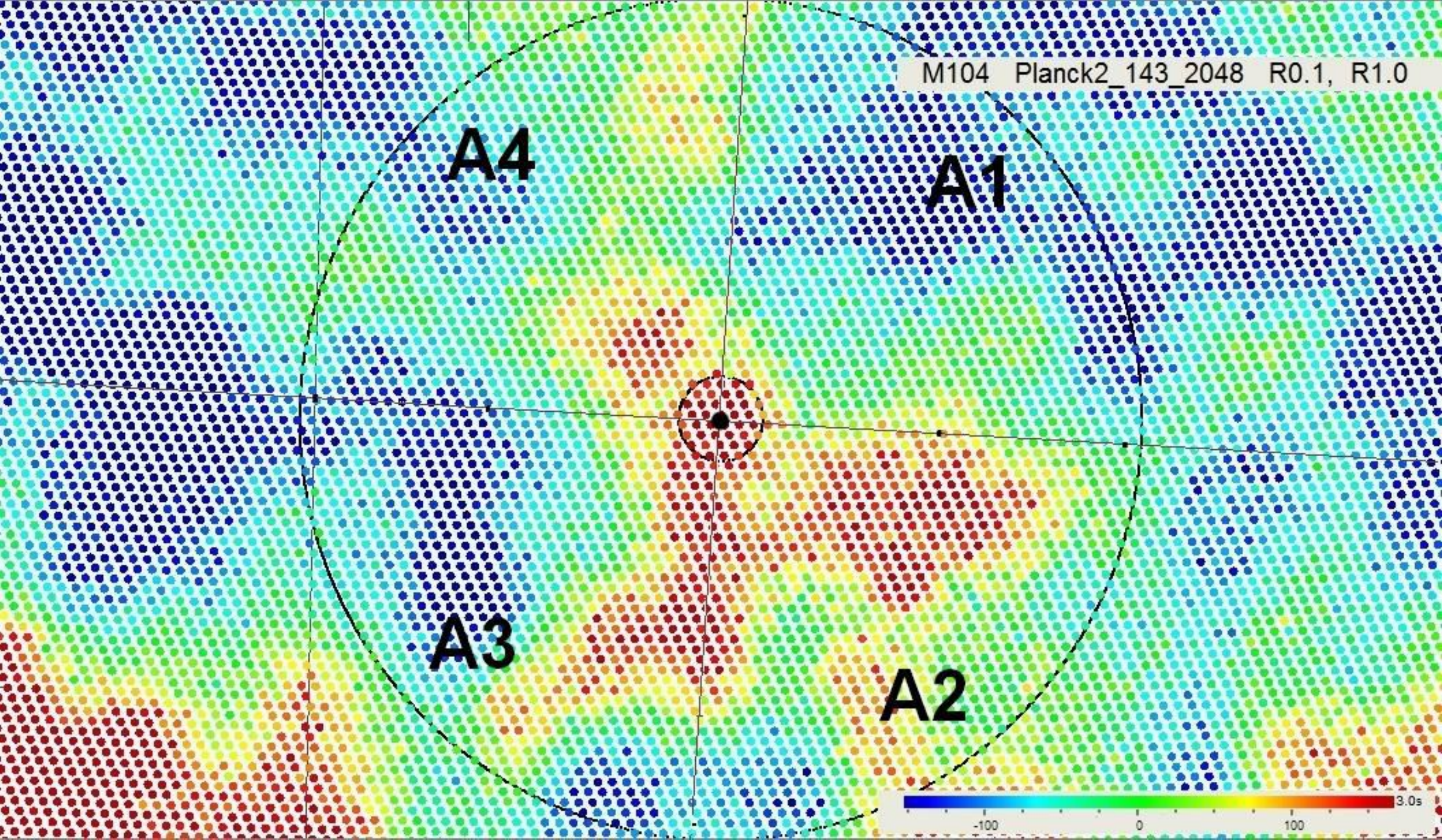}
\caption{ Sombrero galaxy in the 143 GHz Planck band. The inner
and outer circles mark the galactocentric distances of $0.1\degr$ and $1
\degr$, respectively.}
\label{FigSombreroCMB}       
\end{figure}

\begin{figure}[ht]
\centering
\includegraphics[width=0.48\textwidth]{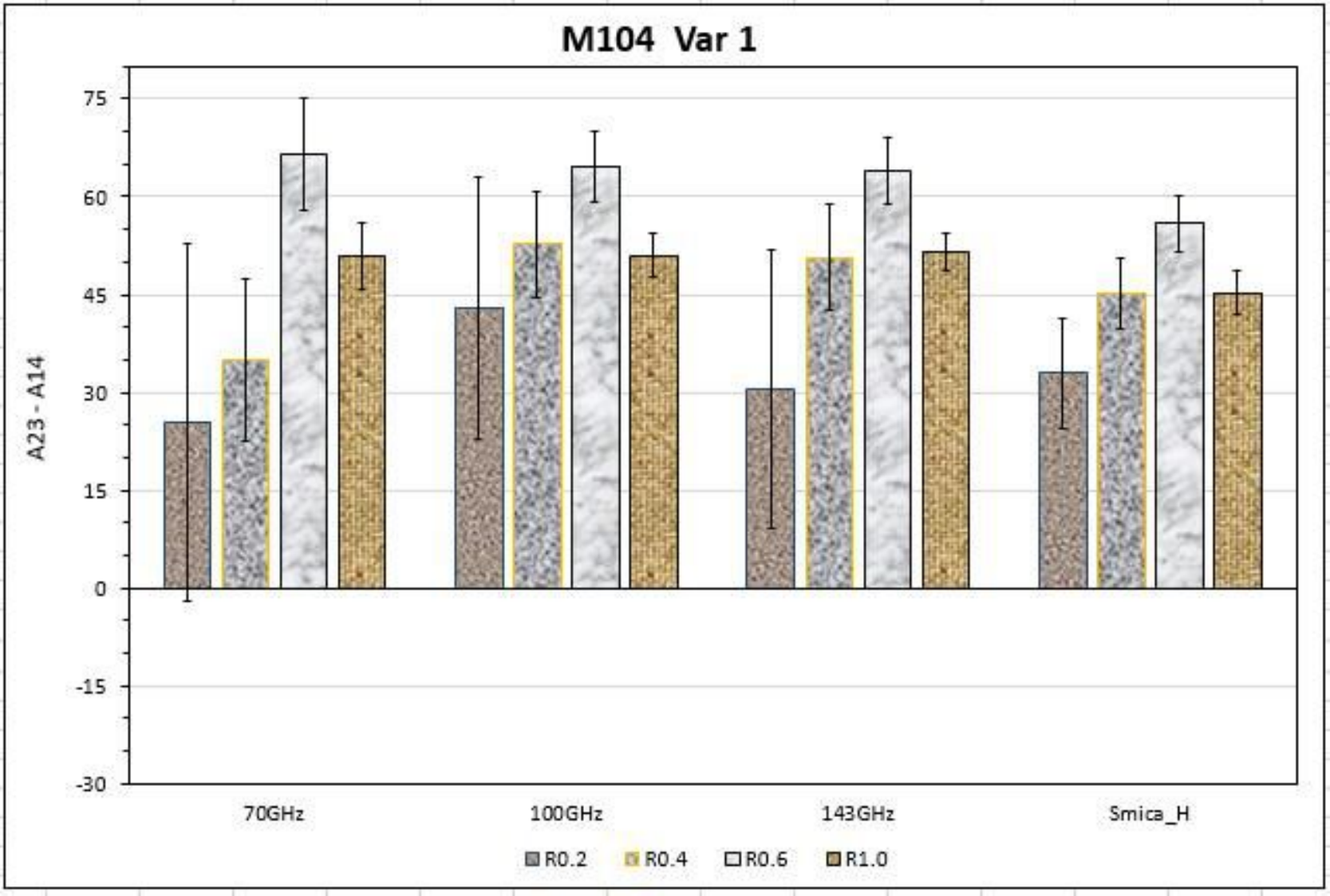}
\includegraphics[width=0.48\textwidth]{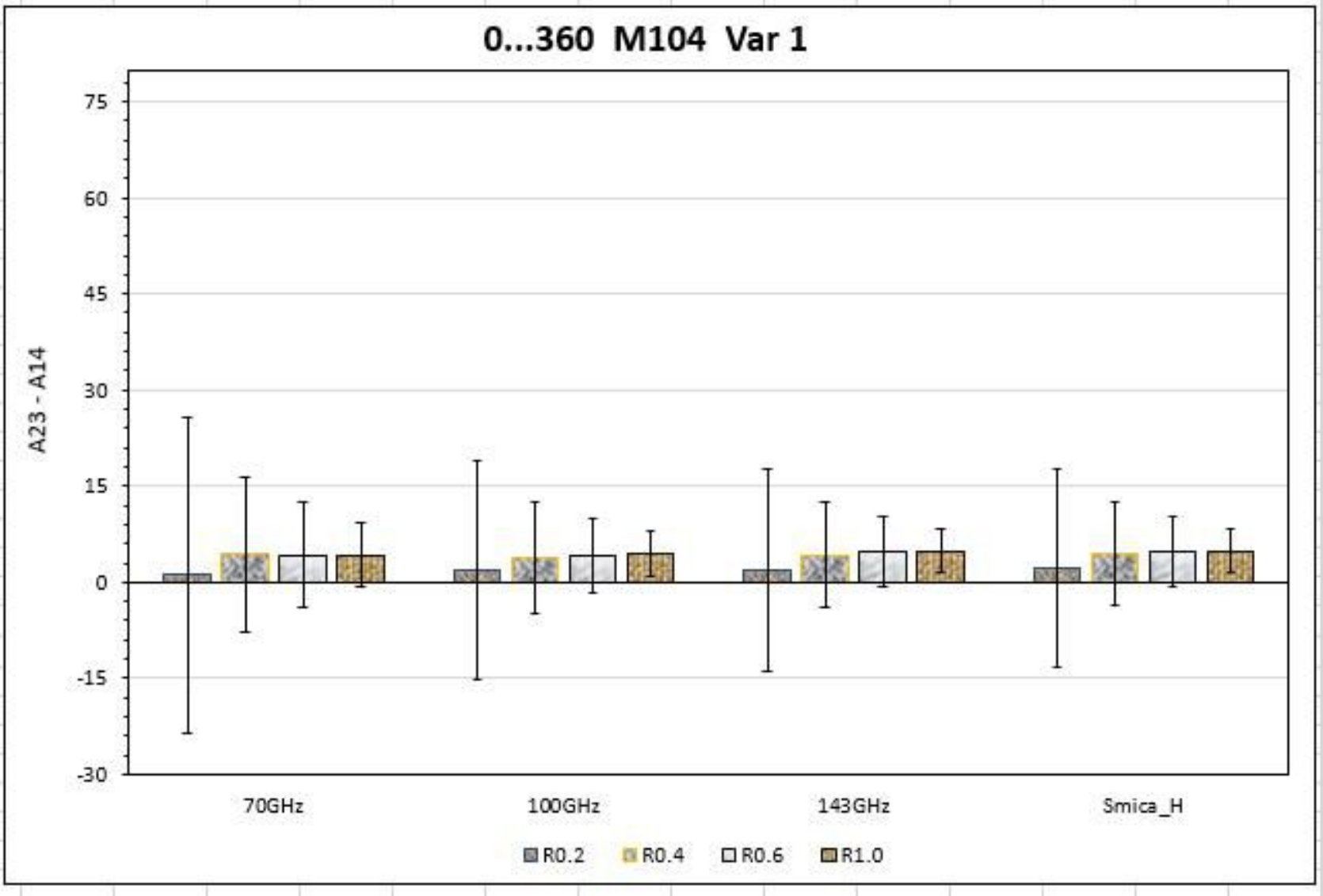}
\caption{Temperature asymmetry towards 
(first panel) the M104 galaxy in $\mu$K (with the standard errors) of the Variant 1 (A2A3 -
A1A4), in the four considered {\it Planck} bands (see text for details)
within four radial distances of  $0.2\degr$ ($R0.2$), $0.4\degr$ ($R0.4$),
$0.6\degr$ ($R0.6$) and $1\degr$ ($R1.0$); and (second panel) the same for the  360 control fields with the same geometry equally spaced at one
degree distance from each other in Galactic longitude and at the same
latitude as M104.}
 \label{fig3}
 \end{figure}

  \begin{figure}[ht]
\centering
\includegraphics[width=0.48\textwidth]{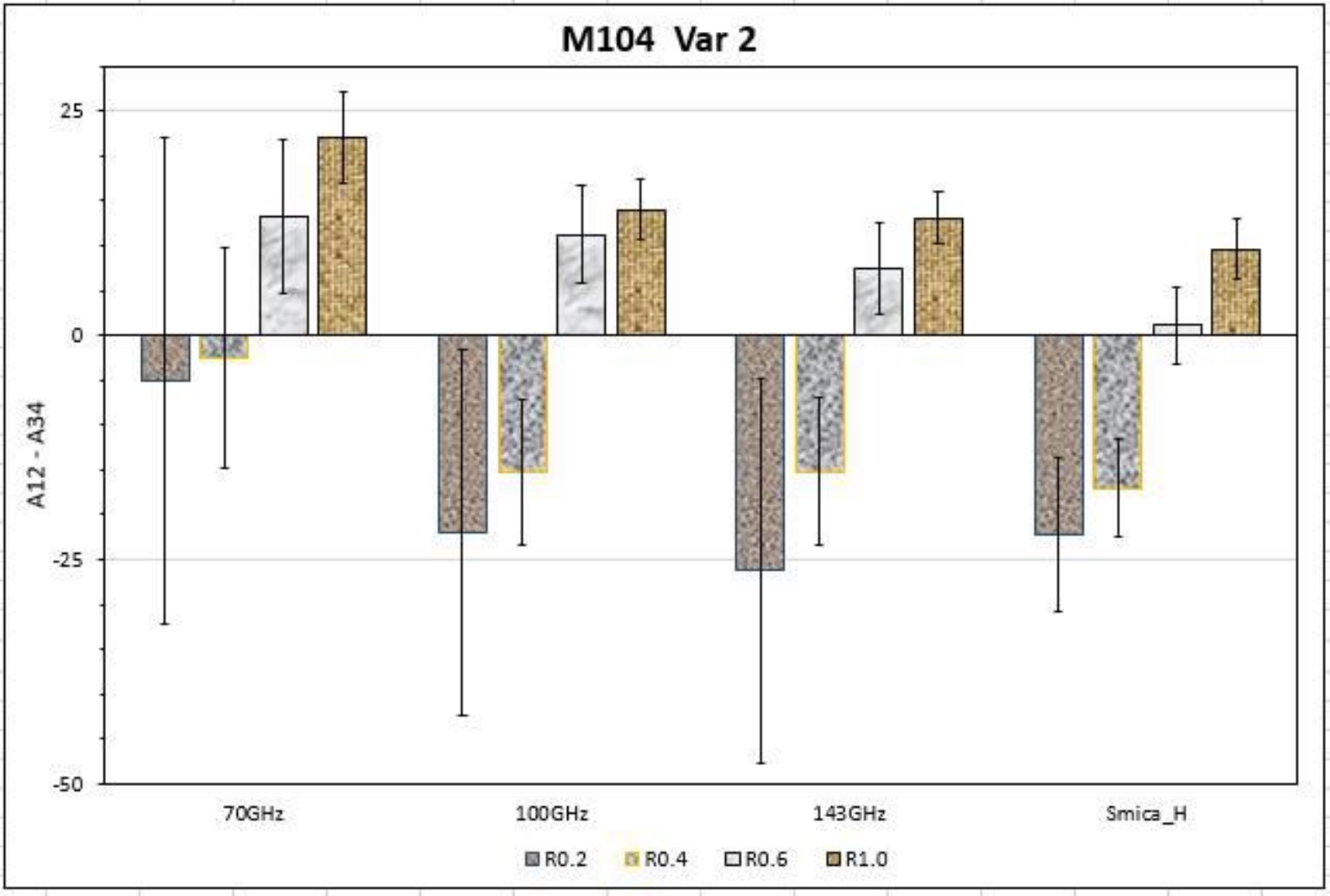}
\includegraphics[width=0.48\textwidth]{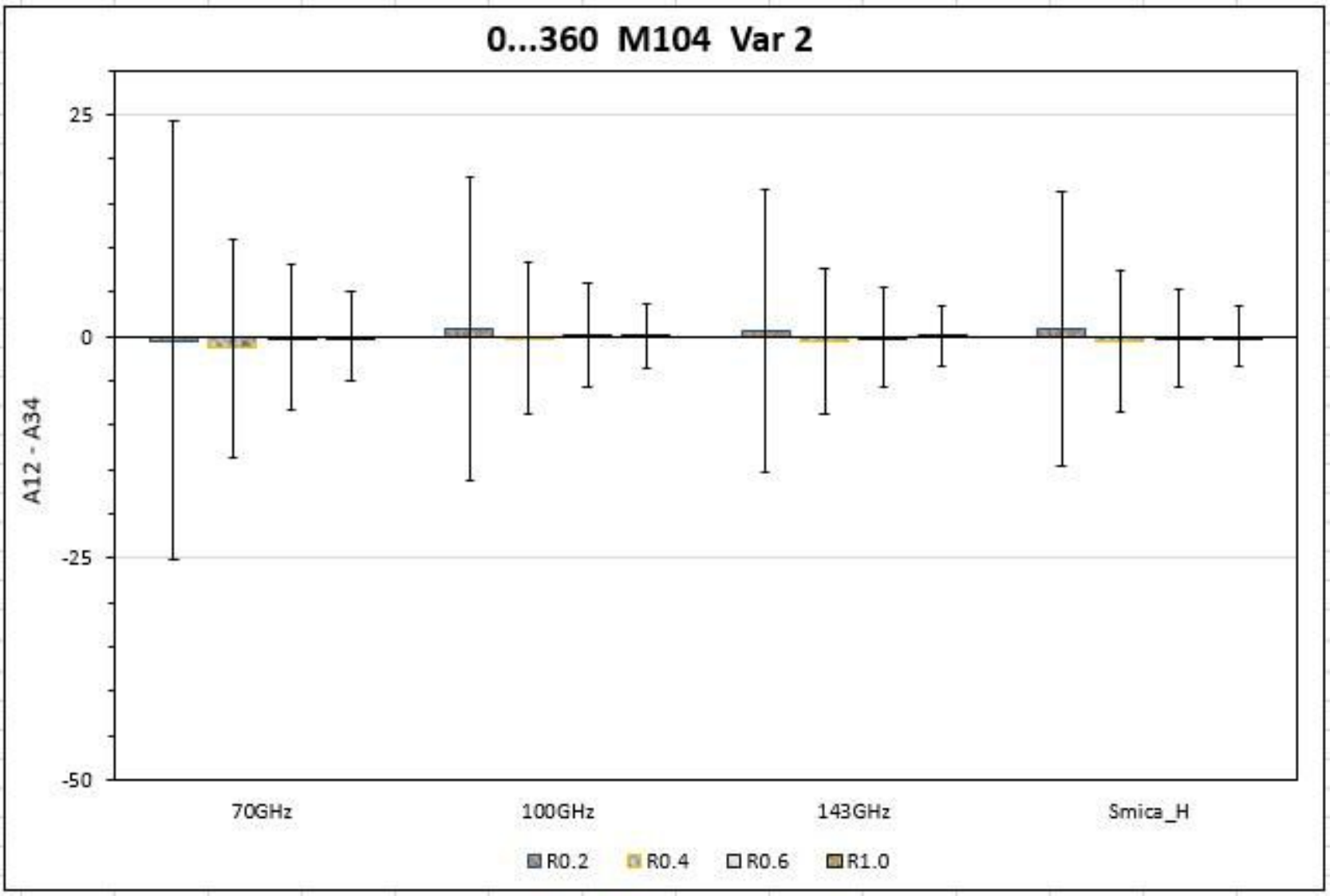}
\caption{Temperature asymmetry toward
(first panel) the M104 galaxy in $\mu$K (with the standard errors) of the Variant 2 (A1A2 -
A3A4), in the four considered {\it Planck} bands (see text for details)
within four radial distances of  $0.2\degr$ ($R0.2$), $0.4\degr$ ($R0.4$),
$0.6\degr$ ($R0.6$) and $1\degr$ ($R1.0$); and (second panel)  the same
for the 360 control fields with the same geometry equally spaced at one
degree distance from each other in Galactic longitude and at the same
latitude as M104.}
 \label{fig4}
 \end{figure}

\begin{figure}[ht]
\centering
\includegraphics[width=0.48\textwidth]{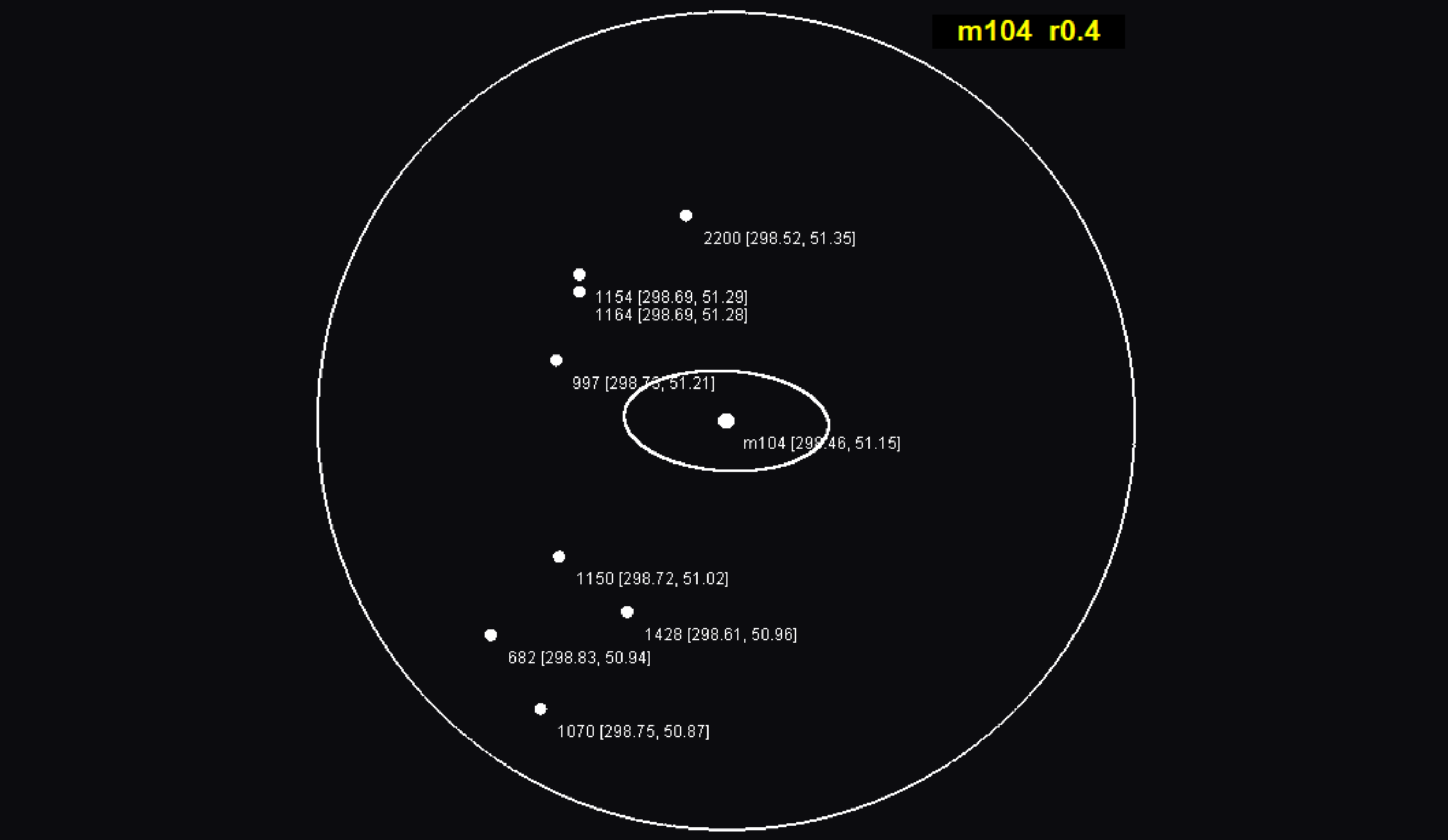}
\caption{Plot of the position of the eight  outermost globular clusters
observed in the M104 galaxy (for details see \cite{bridges2007}). The inner
circle traces the visible part of the Sombrero galaxy while the outer circle
is at  a radius of $0.4\degr$.}
 \label{fig5}
 \end{figure}

\section{The M104 Sombrero galaxy}
\subsection{M104: generalities}
The galaxy M104, also known as the Sombrero galaxy (or NGC 4594), is a majestic galaxy
located in the Virgo constellation at a distance of about 9.55 Mpc from Earth
(see Fig. \ref{FigSombrero} for a view in the optical band). The luminous
bulge and the prominent dust lane  give this galaxy its name.  This dust
lane is the site of star formation in the galaxy and a rich system of
globular clusters (GCs) characterize the M104 halo. The estimated number of
these GCs is about $2\times 10^3$, about ten times larger than the number of
GCs in the Milky Way. It is the brightest nearby spiral galaxy and is
inclined at an angle of only about $7\degr$ to our line of sight and appears
edge-on (its inclination angle is therefore $i=83\degr$). It is visible with
binoculars and small telescopes, but only appears as a small patch of light.
The  large bulge of the Sombrero galaxy and the super-massive black hole at its core
make it a popular target for study. The galaxy, at coordinates RA:
$12^{\rm h}39^{\rm m}59.4^{\rm s}$, Dec: $-11\degr 39\arcmin 23\arcsec$ (the
galactic coordinates are $l=298.46055\degr$ and $b=51.14929\degr$), has major
and minor diameters of about $8.6\arcmin$ and $4.2\arcmin$, corresponding to
about 24 kpc and 9.7 kpc, respectively (at the M104 distance $1\arcmin$
corresponds to about 2.78 kpc and $1\degr$ to about 167 kpc).  The galaxy has
a visible mass  of about $(22.9\pm 3.2)\times 10^{10}$ $M_{\odot}$
\citep{tempel2006} and its  center is thought to be home to a supermassive
black hole.

\subsection{Planck data for M104}
Following the same procedures described in our previous papers  we used
the publicly released {\it Planck} 2015 data \footnote{From the {\it Planck}
Legacy Archive, http://pla.esac.esa.int.} \citep{planck2015a} in the 70 GHz
bands of the Low-Frequency Instrument (LFI), and in the bands at 100 GHz, 143
GHz, and 217 GHz of the High-Frequency Instrument (HFI). We also used
the foreground-corrected SMICA map (indicated as SmicaH), which should
display the lowest contamination by the galactic foreground. We notice here
that the resolution of {\it Planck}  is $13.2\arcmin$, $9.6\arcmin$,  $7.1\arcmin,$
and $5\arcmin$ in terms of FHWM (full width at  half maximum) at 70, 100,
143, and 217 GHz bands, respectively, and the frequency maps by \cite{planck2015b}
are provided in CMB temperature at a resolution corresponding to Nside=2048
in the HEALPix scheme \citep{gorski2005}. We also note that in the CMB maps
we consider, the  monopole and dipole contributions have been removed.

In Fig. \ref{FigSombreroCMB} the Sombrero galaxy in the 143 GHzPlanck band is
shown. The optical extension of the M104 galaxy is  shown, as indicated by
the inner ellipse. Spectroscopic observations in the optical band and radio
observations at 21 cm show that the galaxy disk has an asymptotic rotation
velocity of  $376\pm12$ km s$^{-1}$ \citep{jardel2011}. Also, the globular
cluster system around the M104 galaxy shows a global rotation with an estimated speed of about 100 km s$^{-1}$ \citep{bridges1997}, with the eastern part moving
towards Earth. To study the CMB data toward the M104
galaxy in the simplest way, the {\it Planck} field of the region of  interest  has been divided
into  four quadrants: A1, A2, A3, and A4.

As detailed in the histograms in Figs. \ref{fig3} and \ref{fig4}, we
considered the temperature asymmetry in three radial regions about the M104
center within $0.2\degr$, $0.4\degr$, $0.6\degr$, and  $1.0\degr$ (indicated
as  R0.2, R0.4, R0.6 and R1.0, respectively). In the upper panel of Fig.
\ref{fig3} we give  the temperature asymmetry toward M104 in $\mu$K (with the
standard errors) of the A1+A4 region  with respect to the A2+A3 region in the
four considered {\it Planck} bands within the four radial distances. In the
bottom panel of the figure we give  the same for the 360 control fields with
the same geometry (shown in Fig. \ref{FigSombreroCMB}) equally spaced at one
degree distance from each other in Galactic longitude and at the same
latitude as M104. Here, we mention that we have  intentionally avoided the
use of CMB simulations to get the error bars. Indeed,  simulations are
mandatory when whole-sky CMB properties are studied (correlation functions,
power spectra, etc.), while they could introduce additional uncertainties
related to the modelling of foregrounds. As one can see from Fig. \ref{fig3},
the A1+A4 region always appears hotter than the  A2+A3 region, and the
temperature asymmetry increases with increasing galactocentric radius
from about $25~\mu$K (at  $0.2\degr$) to a maximum value of about  $65~\mu$K
(within $0.6\degr$). We note that SMICA\_H data   show a similar trend.
In the lower panel of   Fig. \ref{fig3}, the 360 control fields show a
temperature asymmetry consistent with zero in all {\it Planck} bands.

In Fig. \ref{fig4} we show the second considered variant, namely the
temperature asymmetry in the four considered {\it Planck} bands of the A3+A4
region with respect to the  A1+A2 one. As one can see by comparing the
upper and lower panels of  Fig. \ref{fig4}, the temperature asymmetry 
now has
a more complicated pattern; it is almost consistent with zero within
$0.4\degr$ in the 70 GHz,  100 GHz, and 143 GHz bands, while SMICA\_H data
indicate that the A1+A2 region is hotter than the A3+A4 one. The A3+A4  side
of the external M104 halo instead seems to be hotter than the other region
in all considered bands by about $10-20 ~\mu$K.

We also considered available data of the GCs in the M104 galaxy.
\cite{bridges2007} searched for rotation of the globular cluster systems
around the M104 galaxy by considering the line of sight velocity of 108 GCs
and found no evidence for global rotation. As also noticed by these latter authors,
the lack of rotation is certainly surprising because there is a rotation of
$300-350$ km s$^{-1}$ in the stellar and gas disk and, moreover, cosmological
simulations of galaxy formation predict a significant amount of angular
momentum in early-type galaxies. In this respect, we note that most of the GCs
towards M104 are detected relatively near to the galactic center. If one
considers the 43  GCs with galactocentric distance larger than $5\arcmin$ one
finds a non-negligible rotation with the $95\%$ upper limit on the rotation
velocity of $250$ km s$^{-1}$ and a GC velocity dispersion $\sigma$ of about
155 km s$^{-1}$. Therefore, the obtained  $95\%$ upper limit on the value of
($v/\sigma$) is about $1.6$.  To further strengthen this discussion, in Fig.
\ref{fig5} we show the position of the eight  outermost GCs (see
\citealt{bridges2007} for details). Unfortunately, these GCs reside in the A3
and A4 regions in Fig. \ref{FigSombreroCMB}, and therefore they can be used at most to
probe the north--south asymmetry (as in the Variant 1), but not the east--west
asymmetry. We obtain that the average line-of-sight velocity of the four GCs in
the A3 region ($<v_r>\simeq 947$ km s$^{-1}$) is similar to that of the four
GCs in the A4 region ($<v_r>\simeq 972$ km s$^{-1}$), also consistent with
the observed recession velocity of the M104 galaxy of about 1024 km s$^{-1}$.
We note here that the two GCs with highest and lowest radial velocity are
GC2200 in the A4 region and GC682 in the A3 region. These two globular
clusters show a remarkable rotation about the M104 center since the first one
is moving with  line-of-sight  velocity  $<v_r>\simeq 1524$ km s$^{-1}$ while
GC682 is moving towards Earth with  ($<v_r>\simeq 681$ km s$^{-1}$), thus
with a relative velocity of  $\simeq 343$ km s$^{-1}$ towards the observer.
There is no doubt that an enlarged sample of GCs in the M104 galaxy at
galactocentric distances larger than $0.4\degr$ could greatly help to probe the
rotation of the Sombrero galaxy  halo.

The detected temperature structure observed in the {\it Planck} microwave
bands towards  M104 may indicate the absence of a regular bulk motion at
various scales  and  complicated internal dynamics. Indeed, the presence of
an internal disk  and an external ellipsoid has been suggested to explain the
complex dynamics of the M104 galaxy.

We would like to emphasize that our methodology to study the temperature
asymmetry towards a given target aims to be maximally model-independent and,
in particular, independent of the precise knowledge of the spectrum emitted
in the considered regions (due to molecular clouds or to other mechanisms
such as the kinetic Sunyaev-Zel'dovich effect  or synchrotron emission,
etc.). The temperature asymmetry signal we detect is practically frequency
independent \footnote{We note that the slight apparent anomalies at small
radii, where error boxes are larger, are due to the low number of pixels.},
which is a clear signature of the Doppler effect.  In fact, we find that, in
the case of the first Variant, the probability that the detected signal in
each of the four considered rings is due to a random fluctuation of the CMB
signal is $0.30, 0.31, 0.11, 0.26$, respectively,  which yield, assuming
independent probabilities, a cumulative probability of about $1.8\times
10^{-3}$, that is less than $0.2\%$, while for Variant 2, where the situation
is less regular, the chance probability yields $3.1\times 10^{-2}$.

Before closing this section, we also note that the simple Doppler nature of
the temperature asymmetry can be influenced by several effects such as
the peculiar motion of the halo clouds, the nonuniform spectrum of the clouds due to
the radiation balance and radiative transfer details, and so on. However,
obviously, at this stage, those are lower-order effects with respect the
global halo rotation. Moreover, since we consider M104 as being foreground to the CMB, we
note that for the maps we study (with monopole and dipole extracted) the
foregrounds could also be negative due to physical processes, for example, in the
case of clouds, the particular molecular energy level transitions, metastable
states, two-photon emission or absorption, and so on.

\section{Conclusions}
Similar to the case of  other galaxies of the Local Group considered
previously  (in particular the galaxies M31, M82, M81, M33, and Cen A have been analyzed using {\it Planck} data and a temperature asymmetry $\Delta T$ about $60-80$ $\mu$K  of one side with respect to the other about the rotation axis of the galactic disks has  been detected in all cases), we found a consistent north--south temperature asymmetry  toward
the M104 galaxy, that reaches values up to  about $65~\mu$K within $0.6\degr$
(about 100 kpc) in all considered {\it Planck} bands \footnote{We remark here that we consider maps of Nside=2048 in the HEALPix scheme, which corresponds to a pixel size $\simeq 1.718\arcmin$ (\citealt{gorski2005}, see also the web page $https://irsa.ipac.caltech.edu/data/Planck/release_1/ancillary-data/$),
thus ensuring the pixel number statistics in the studied regions and the resulting error bars as in Figs. \ref{fig3} and \ref{fig4}, and the discussion in Section 2.2.}, and then starts to
decrease. We mention that Chandra observations in the X-ray band of the M104
galaxy show the presence of a diffuse X-ray emission extending at least up to
30 kpc from its center \citep{Li2011}, together with hundreds of point
sources which are a mixture of objects in the M104 halo (mainly X-ray
binaries) as well as background Quasars \citep{Li2010}. However, the presence of this
X-ray-emitting hot gas cannot explain the temperature asymmetry
detected in {\it Planck} data.

It is straightforward to show that if the cold-gas cloud model is at the origin of the
detected temperature asymmetry (however, as anticipated in the previous
section, other emission mechanisms may play a role in the detected
signal), a lower limit to the M104 galaxy dynamical mass is given by
\begin{equation}
M_{\rm dyn}(R)\simeq 700 M_{\odot}\left(\frac{R}{100\, 
{\rm kpc}}\right)\frac{\Delta T^2_{\mu k}}{(\tau_{\rm eff}\sin i)^2},
\end{equation}
where $R$ is the considered galaxy halo radius in units of 100 kpc, $\Delta
T$ is the measured temperature asymmetry (in $\mu k$), and $\tau_{\rm eff}$ is
the effective cloud optical depth (which depends on both the cloud filling
factor and the averaged optical depth within a given {\it Planck} band.
Expected values for $\tau_{\rm eff}$ are about a few $10^{-3}$
\citep{tahir2019}, meaning that  the M104 dynamical mass out to $\sim
100$ kpc is seen to be $M_{\rm dyn}\simeq 3\times 10^{12}   M_{\odot}$, in agreement
with other measurements (see, e.g., \citealt{tempel2006}). It is interesting to
mention in this respect that this galactocentric distance appears close to
the transonic point (at $\simeq 126$ kpc), determined recently  by
\cite{igarash2014} within the framework of  the transonic outflow model in a
dark matter halo applied to the Sombrero galaxy. \footnote{Indeed, outflow
could be among the gas-heating mechanisms, together with shocks, turbulence, and
galactic merging. The outflow itself could be uncorrelated to the 
rotation of the halo but the heated gas can contribute to the temperature
asymmetry.}

The detected temperature asymmetry is, as discussed in the previous section,
almost frequency-independent and indicates  that the M104
galaxy halo is rotating with respect to  the major symmetry axis of the galaxy disk
(variant 1). However, the fact that a strong temperature asymmetry is
detected  also with respect to the minor symmetry axis of M104 (variant 2),
is a robust indication of a relatively complex geometry and rotation patterns. We
note that   that several
different components have been shown to contribute to the observed galaxy kinematics (see, e.g., \citealt{wagner1989}). These
components have different rotational velocities and velocity dispersions. In
addition, they cannot be described by a single Gaussian broadening function.
This complex structure may explain the observations in the different {\it
Planck} bands toward the Sombrero galaxy. As a matter of fact, the Sombrero
galaxy exhibits the characteristics of both disk-like and elliptical
galaxies. Recent observations by the Spitzer space telescope show that M104 is
much more complex than previously thought and resembles  a disk galaxy
inside an elliptical one. We wish to emphasize here that  the results in the
present paper find strong support by Spitzer infrared vision  which offers  a
different view of that emerging in visible light, with a glowing halo filled
by old stars and a strong quantity  of dust through which we observe a
reddened star population. This implies that the M104 halo has the same size
and mass as those of giant elliptical galaxies. The
Sombrero galaxy resembles a giant elliptical that has swallowed a disk galaxy,
but this is unlikely to have taken place since the process would have completely destroyed the
galactic disk structure.  Following \cite{gadotti2012}, we propose that the
M104 galactic structure is the result of a giant elliptical inundated by a
huge amount of gas billions of years ago. This could explain the complex
dynamical morphology of the Sombrero galaxy.

Before concluding this paper, we would like to make a general remark about the
use of CMB data to map various galaxy halos. Indeed,  the systematic study of
the mean temperature asymmetry, applied here to the Sombrero galaxy, could
become a  conventional tool for studying and mapping the internal motions
of not-too-distant galaxies (including their halos) in the microwave band,
in a complementary way with respect to other methods. As for the case of the
Kolmogorov stochasticity parameter (\citealt{gurzadyan2009,gurzadyan2014})
and the Sunyaev-Zeldovich effect, software for an automated analysis,
cross-correlating surveys in different bands with CMB data, may be developed.
This is especially important in view of the  next generation of CMB experiments,
such as LiteBird \citep{LiteBird}, CMB-S4 \citep{CMB-S4}, CORE (Cosmic
Origins Explorer, \citealt{CORE}), DeepSpace\footnote{See the DeepSpace
website at http://deep-space.nbi.ku.dk.}, PIXIE (Cosmic Origins Explorer,
\citealt{PIXIE}),  and Polarbear \citep{Polarbear}, which will attempt to obtain
 measurements of the CMB that are even more precise than have been available so far and that may allow higher-resolution studies of the galaxies of the Local Group. 
Many of these experiments are designed to cover mainly the frequency range around 100
GHz where the relative intensity of the CMB is known to be highest and where
one of the most dominant foreground components is dust emission (see, e.g.,
\citealt{liu2017}). Understanding the properties of dust emission and
distinguishing between Galactic foregrounds and extragalactic emission are 
important for the optimized use of the next-generation CMB
experiments.

\begin{acknowledgements}
We acknowledge the use of {\it Planck} data in the Legacy Archive for
Microwave Background Data Analysis (LAMBDA) and HEALPix \citep{gorski2005}
package. FDP, GI and AAN acknowledge the TAsP and Euclid INFN projects. PJ
acknowledges support from the Swiss National Science Foundation. AQ is most grateful to the Lecce Unit of INFN for support during this work.
\end{acknowledgements}


\end{document}